%% file: main_AAAI24.tex
\documentclass[letterpaper]{article} 
\usepackage{aaai24}  
\nocopyright 
\usepackage{times}  
\usepackage{helvet}  
\usepackage{courier}  
\usepackage[hyphens]{url}  
\usepackage{graphicx} 
\usepackage{natbib}  
\usepackage{caption} 
\usepackage{amsmath,amssymb}
\usepackage{amsthm}
\usepackage{xcolor}
\usepackage{newfloat}
\usepackage{listings}
\usepackage{diagbox}
\usepackage{algorithm}
\usepackage{algpseudocode}
\usepackage{amsthm}
\usepackage[english]{babel}
\usepackage{hhline}
\usepackage{subcaption}
\usepackage{enumitem}
\usepackage{graphicx}

\theoremstyle{remark}

\usepackage{microtype}
\usepackage{graphicx}
\usepackage{booktabs} 
\usepackage{amssymb}
\usepackage{multirow}
\usepackage[textsize=tiny, textwidth=1.1cm]{todonotes}

\usepackage{tikz}
\usepackage{times}
\usepackage{pbox}
\usepackage{mathtools} 
\usepackage{enumitem}
\usepackage{amsxtra}

\DeclareUnicodeCharacter{2212}{-}

\usepackage{xcolor}
\usepackage{comment}

\usepackage{newfloat}
\usepackage{listings}
\DeclareCaptionStyle{ruled}{labelfont=normalfont,labelsep=colon,strut=off} 
\floatstyle{ruled}
\newfloat{listing}{tb}{lst}{}
\floatname{listing}{Listing}
\title{Real-time Network Intrusion Detection via Decision Transformers} 
\author {
    Jingdi Chen\textsuperscript{\rm 1},
    Hanhan Zhou\textsuperscript{\rm 1},
    Yongsheng Mei\textsuperscript{\rm 1},
    Gina Adam\textsuperscript{\rm 1}
    Nathaniel D. Bastian\textsuperscript{\rm 2},
    Tian Lan\textsuperscript{\rm 1}, 
}
\affiliations {
    \textsuperscript{\rm 1}George Washington University, Washington, DC\\
    \textsuperscript{\rm 2}Army Cyber Institute, United States Military Academy, West Point, NY\\
    jingdic@gwu.edu, hanhan@gwu.edu, ysmei@gwu.edu, ginaadam@gwu.edu,  nathaniel.bastian@westpoint.edu, tlan@gwu.edu
}

\begin{document}

\maketitle






\begin{abstract}


Many cybersecurity problems that require real-time decision-making based on temporal observations can be abstracted as a sequence modeling problem, e.g., network intrusion detection from a sequence of arriving packets. Existing approaches like reinforcement learning may not be suitable for such cybersecurity decision problems, since the Markovian property may not necessarily hold and the underlying network states are often not observable. In this paper, we cast the problem of real-time network intrusion detection as casual sequence modeling and draw upon the power of the transformer architecture for real-time decision-making. By conditioning a causal decision transformer on past trajectories, consisting of the rewards, network packets, and detection decisions, our proposed framework will generate future detection decisions to achieve the desired return. It enables decision transformers to be applied to real-time network intrusion detection, as well as a novel tradeoff between the accuracy and timeliness of detection.
The proposed solution is evaluated on public network intrusion detection datasets and outperforms several baseline algorithms using reinforcement learning and sequence modeling, in terms of detection accuracy and timeliness.

\end{abstract}

\input{1Intro}

\input{2RelatedWork}

\input{3Preliminary}

\input{4ProblemFormulation}

\input{5Methodology}

\input{6Experiment}

\input{7Conclusion}

\section*{Acknowledgements}

This work was supported in part by the U.S. Military Academy (USMA) under Cooperative Agreement No. W911NF-22-2-0089. The views and conclusions expressed in this paper are those of the authors and do not reflect the official policy or position of the U.S. Military Academy, U.S. Army, U.S. Department of Defense, or U.S. Government.

\bigskip
\bibliography{aaai24}

\end{document}

%% file: 1Intro.tex
\section{Introduction}

Machine learning has been successfully applied to many network security problems, such as packet inspection and network intrusion detection~\cite{6994301}, which can be formulated as real-time decision-making problems based on temporal observations (e.g., a sequence of packets or network activities). Existing approaches such as Reinforcement Learning (RL) leverage value functions or policy gradients for training~\cite{zhou2022pac,zhou2023value,chen2023minimizing}. In addition to high complexity and computational overhead, RL-based approaches~\cite{mei2023remix,mei2023mac,gogineni2023accmer} may not be suitable for cybersecurity decision-making problems, where the Markovian property may not necessarily hold and the underlying network states are not observable. To this end, 
consequently, there is a critical need for a simple and scalable solution, yet capable of 
identifying attacks/intrusions at the packet level to achieve accelerated detection speed. 
Further, it should enable timely detection of malicious packets, potentially intercepting them before the completion of the entire network flow. 
It would not only enhance the timeliness of the detection process but also significantly improve the overall efficiency and effectiveness of network defense.
To address these challenges, we develop a framework to abstract network intrusion detection as a sequence modeling problem for decision-making. It allows us to leverage the powerful decision transformer architecture~\cite{dt}, as well as associated advances in language modeling such as GPT-x and BERT. Specifically, our proposed method will feed past trajectories, consisting of the rewards, network packets, and detection decisions into a causally masked \textbf{decision transformer (DT)}, to generate future detection decisions (including waiting for more observations, flagging as malicious, and flagging as benign), in an autoregressive fashion to achieve the desired return. It is 
simple and elegant, extends the success of transformer models in sequence processing, as seen in language models~\cite{vaswani2017attention}, to sequence modeling and decision-making. 
We note that such sequence modeling is especially relevant for network intrusion detection, where malicious activities may occur at any point in packet transmission. This trajectory-focused approach offers more precise detection than traditional transition-based RL methods. Notably, DT does not hinge on the Markov property (it refers to the property of a system where the probability of future states depends only on the current state and not on the sequence of events that preceded it, in other words, the future is independent of the past given the present), a common yet restrictive assumption in Markov decision process (MDP)-based RL methods~\cite{chen2021bringing,zhou2022federated,chen2023distributional} that may not always hold in network intrusion contexts. Furthermore, DT's sequence modeling transcends the need for time-homogeneous transitions, introducing greater flexibility for intrusion detection applications.
It's important to note that while DT primarily uses transformer architecture for its sequence model, its training is fundamentally architecture-agnostic. This flexibility allows for the use of alternative deep autoregressive models, such as Long Short Term Memory (LSTM) networks~\cite{hochreiter1997long} or Temporal Convolutional Networks (TCNs)~\cite{bai2018empirical}. Parallel to DT, the Trajectory Transformer~\cite{janner2021offline} also employs a similar sequence modeling approach with a transformer architecture but incorporates model-based planning. This necessitates discretizing states and actions and predicting future states and returns.

\textbf{Our Contribution. }  
In this paper, we formulate network intrusion detection as a sequence modeling problem and investigate the effectiveness of DT in network intrusion detection applications. More precisely, we consider a set of trajectories, each made up of a sequence of network packets/states, detection labels/actions, and target design rewards, and focus on learning meaningful patterns (as represented by a decision-making policy) from collected trajectories that are \textbf{not necessarily optimal}. The learned policy then generates future detection actions at testing time by conditioning on an autoregressive model of the desired return, past network packets/states, and previous actions. 
For better incorporation of the packet-level features, the proposed solution makes novel use of an Autoencoder~\cite{zhai2018autoencoder} to efficiently compress arbitrary-length packet sequences into more compact packet embeddings for input features of the DT. 
Then, we investigate a novel tradeoff between detection \textbf{accuracy} and detection \textbf{timeliness} by introducing a reward function that penalizes any delayed (yet possibly more accurate due to more available observations) detection decisions.
The recent study~\cite{brandfonbrener2022does} underscores a pivotal limitation of the DT framework; its dependency on the quality of the behavior policy in the training data. This concern, however, is substantially alleviated in the domain of network intrusion detection, where datasets such as UNSW-NB15\cite{moustafa2015unsw} inherently provide high-quality data, distinctly classified into malicious and benign records. Such well-differentiated datasets are conducive to training DT, ensuring it learns effectively from representative examples of network behaviors. Furthermore, we propose to enhance DT’s applicability by incorporating techniques like importance sampling, which strategically weighs different network traces, thereby refining the model's performance in accurately detecting network intrusions. This approach aligns well with the advanced analytical needs of top-tier cybersecurity frameworks, offering a nuanced solution to the challenge of effective intrusion detection. 
We highlight some notable features of our proposed framework:
\begin{itemize}
    \item We propose a novel algorithm leveraging an autoencoder to integrate packet payload features into compressed embeddings that serve as packet-level, sequential input suitable for a causal transformer architecture.
    \item  We formulate the network intrusion detection (NID) problem as sequence modeling for decision-making and leverage a decision transformer architecture to explore a new tradeoff between accuracy and timeliness in network intrusion detection problems.
   \item We investigate limitations related to learning-based NID algorithms and highlight DT’s capability in NID by testing it with non-optimal collected trajectories through different sampling techniques and evaluating the proposed solutions on real-world datasets.
\end{itemize}

%% file: 2RelatedWork.tex
\section{Related Work}

\textbf{Learning-based NIDS. }
Network intrusion detection systems (NIDS) have become increasingly important in network security, and various machine learning methods, including supervised, semi-supervised, and unsupervised learning, have been used to enhance their accuracy and precision in detecting anomalies. Several studies have proposed NIDS for IoT networks using different machine learning algorithms and architectures, such as Multilayer Perceptron (MLP)~\cite{hodo2016threat}, Artificial Immune System (AIS)~\cite{hosseinpour2016intrusion}, 
Internet of Things Intrusion Detection and Mitigation (IoT-IDM)~\cite{nobakht2016host}, 
and Conditional Variational Autoencoder (CVAE)~\cite{lopez2017conditional}. Other studies have recommended the use of fog computing to improve efficiency and scalability in IoT systems~\cite{diro2018distributed}.
Previous Deep Learning-based approaches in NIDS exhibit certain limitations, notably their ineffectiveness in assimilating features from new packets as they first appear. This shortfall impedes the ability to achieve real-time detection. Additionally, many of these NIDS algorithms operate under the assumption of the Markov property, which may not always hold true in practical scenarios. The nature of network traffic, especially the sequential patterns of packet data, necessitates a more nuanced analysis for enhancing both the speed and accuracy of intrusion detection. This complexity often extends beyond the scope of what can be adequately addressed by the Markovian assumption alone.

\textbf{Offline RL for NIDS. }
The exploration of Deep Reinforcement Learning (DRL) in the realm of NIDS is garnering significant interest, as highlighted in recent studies~\cite{nguyen2021deep}.
Besides, \citet{ren2022id} presented a network intrusion detection model ID-RDRL based on RFE feature extraction and deep reinforcement learning, filtering the optimum subset of features using the RFE feature selection technique with DRL to recognize network intrusions.
\citet{Ren2023MAFSIDSAR} proposed an RL-based intrusion detection model with multi-agent feature selection networks, MAFSIDS, utilizing a feature self-selection algorithm and DRL for intrusion detection.
Traditionally aligned with online environments, DRL operates within the framework of a Markov Decision Process (MDP) suited for interactive settings. This approach, however, is not directly applicable to scenarios involving pre-recorded attack datasets, a domain where offline reinforcement learning becomes pertinent~\cite{levine2020offline}. Despite its relevance, applications of offline reinforcement learning in network intrusion detection remain relatively sparse, with only a few notable instances~\cite{lopez2020application,caminero2019adversarial}. Additionally, none of these studies have demonstrated the capability to detect attacks at the packet level. This gap underscores the significance of offline reinforcement learning as a burgeoning field of research and a potential alternative or extension to traditional supervised learning methodologies in NIDS~\cite{levine2020offline}.

%% file: 3Preliminary.tex
\section{Preliminaries}

\textbf{Offline Reinforcement Learning. }We adopt the Markov decision process (MDP) framework to model our environment, denoted as \(\mathcal{M} = (\mathcal{S}, \mathcal{A}, p, P, R, \gamma)\), where \(\mathcal{S}\) represents the state space, \(\mathcal{A}\) denotes the action space, \(p(s_{t+1} | s_t, a_t)\) is the probability distribution over transitions, \(R(s_t, a_t)\) stands for the reward function, and \(\gamma\) represents the discount factor. At the outset, the agent commences in an initial state \(s_1\) sampled from \(p(s_1)\). Subsequently, at each timestep \(t\), it selects an action \(a_t\) from \(\mathcal{A}\) while in state \(s_t \in \mathcal{S}\), transitioning to \(s_{t+1}\) according to \(p(s_{t+1} | s_t, a_t)\). Following each action, the agent receives a deterministic reward \(r_t = R(s_t, a_t)\). The primary objective in reinforcement learning is to learn a policy that maximizes the expected return, \(E[\sum_{t=1}^{T} r_t]\), within the MDP. In the context of offline reinforcement learning, instead of obtaining data via environment interactions, we only have access to a fixed and limited dataset comprised of trajectories sampled from the environment. This paradigm shift eliminates the agent's capacity to engage in active exploration of the environment, posing a more challenging learning scenario.

\textbf{Decision Transformer. }The Decision Transformer~\cite{dt,chen2023contiformer,zhang2023continuous} processes a trajectory \(\tau\) as a sequence of three types of input tokens: Return-to-Go (RTGs), states, and actions, represented as \(\hat{R}_1, s_1, a_1, \hat{R}_2, s_2, a_2, \ldots, \hat{R}_{|\tau|}, s_{|\tau|}, a_{|\tau|}\). Herein, we use the symbol \(\tau\) to represent a trajectory, and \(|\tau|\) indicates the length of this trajectory. The concept of RTG for a trajectory \(\tau\) at timestep \(t\) is defined as \(\hat{R}_t = \sum_{k=t}^{|\tau|} r_k\), signifying the cumulative sum of future rewards from that particular timestep. We employ \(\boldsymbol{a}\) to denote the sequence of actions, \(\boldsymbol{s}\) to represent the sequence of states, and \(\hat{\boldsymbol{R}}\) to represent the sequence of RTGs associated with the trajectory \(\tau\). Specifically, the initial RTG \(\hat{R}_1\) is equal to the return of the entire trajectory. At timestep \(t\), DT utilizes tokens from the latest \(K\) timesteps to generate an action \(a_t\), where \(K\) serves as a hyperparameter and is also known as the context length for the transformer. It is worth noting that the context length used during evaluation may be shorter than the one employed during training, as we will demonstrate in our experiments. DT learns a deterministic policy \(\pi_{\text{DT}}(a_t | \boldsymbol{s}_{-K,k}, \hat{\boldsymbol{R}}_{-K,k})\), where \(\boldsymbol{s}_{-K,t}\) is a shorthand representation of the sequence of \(K\) past states \(\boldsymbol{s}_{[\text{max}\{1,t-K+1\} : t]}\), and similarly for \(\hat{\boldsymbol{R}}_{-K,t}\). This policy is an autoregressive model of order \(K\). In particular, DT parameterizes the policy using a GPT architecture~\cite{radford2018improving}, which applies a causal mask to enforce the autoregressive structure in the predicted action sequence.

Operating within the DT means that the agent operates within a dynamic training data distribution denoted as \(\mathcal{D}\). Initially, during the pretraining phase, \(\mathcal{D}\) aligns with the offline data distribution \(\mathcal{D}_{\text{offline}}\) and is accessed through an offline dataset denoted as \(\mathcal{D}_{\text{offline}}\). 
For the sake of simplicity, we make the assumption that the data distribution \(\mathcal{D}\) generates associated length-\(K\) subsequences of actions, states, and RTGs, all originating from the same trajectory. To streamline our presentation, we employ a slight abuse of notation and use \((\boldsymbol{a}, \boldsymbol{s}, \hat{\boldsymbol{R}})\) to collectively denote a sample from \(\mathcal{D}\). This notation simplifies the exposition of the training objective for our approach, and the previously defined symbols readily apply in this context. The policy is trained to predict the action tokens using the standard \(\ell_2\) loss defined as:
\[
\mathbb{E}_{(\boldsymbol{a}, \boldsymbol{s}, \hat{\boldsymbol{R}}) \sim \mathcal{D}} \left[\frac{1}{K} \sum_{k=1}^{K} \left(a_k - \pi_{\text{DT}}(\boldsymbol{s}_{-K,k}, \hat{\boldsymbol{R}}_{-K,k})\right)\right]^2. \tag{1}
\]

In practice, we employ uniform sampling to extract length-\(K\) subsequences from the offline dataset \(\mathcal{D}_{\text{offline}}\). During the evaluation phase, we specify the desired performance \(\hat{\boldsymbol{R}}_1\) and an initial state \(s_1\). DT then generates the initial action \(a_1 = \pi_{\text{DT}}(s_1, \hat{R}_1)\). Subsequently, upon generating action \(a_t\), we execute it and observe the resulting next state \(s_{t+1} \sim P(s_{t+1} | s_t, a_t)\), obtaining a reward \(r_t = R(s_t, a_t)\). This process enables us to calculate the next RTG as \(\hat{R}_{t+1} = \hat{R}_t - r_t\). Following this, DT generates the subsequent action \(a_{t+1}\) based on the states \(s_1, s_2\) and the RTGs \(\hat{R}_1, \hat{R}_2\) up to that point in the trajectory.

%% file: 4ProblemFormulation.tex
\section{Problem Formulation}
\label{sec:problem_formulation}

\textbf{Optimization Objectives. }
We aim to enhance network security by optimizing the inspection and response protocols for each data flow in our packet-level NIDS. Technically, each data flow is represented as a stream of time-stamped packet inspections \(\{(t_i, p_i, d_i, w_i)\}_{i=1}^{I_{n}}\): for the \(i^{th}\) packet in flow $F_{n}, n = 1,\dots, N$, where each flow has $I_{n}$ packets, note that each flow has varying numbers of packets $I_{n}$. \(t_i\) marks the time of inspection, \(p_i\) is the set of packet characteristics observed (like protocol and payload), \(d_i\) is the decision taken regarding the packet's nature (benign or malicious), and \(w_i\) is the waiting period before the next packet inspection. Particularly, \(w_i\) represents the elapsed time before inspecting the next packet in the flow, making the next inspection time \(t_{i+1} = t_i + w_i\). This approach focuses on finding the optimal balance between immediate decision-making for rapid threat detection and waiting for subsequent packets to enhance accuracy, creating a novel trade-off in NIDS.

Suppose the instantaneous security status of a network at time \(t\) is represented by \(s(t)\), which may depend on the entire history of packet inspections and responses. Our goal is to maximize the overall network security, represented by \( \int_{0}^{\infty} s(t)dt \). That is, we aim to learn a policy \(\pi\) from which the appropriate decisions \(d_i\) and waiting periods \(w_i\) can be selected for each packet inspection \(i\) in each network flow, so as to maximize the expected total security status, which we refer to as the total security return:
\begin{equation}\label{eq:1}
    \mathrm{max}_{\pi} \mathbb{E}_{\pi} \int_{0}^{\infty} s(t) dt = \mathrm{max}_{\pi}\sum_{n=1}^{N}\sum_{i=1}^{I_{n}}\mathbb{E}_{d_i,w_i \sim \pi}\int_{t_i}^{t_{i+1}} s(t)dt.
\end{equation}
 This involves strategically balancing rapid response to potential threats with the accuracy gained from inspecting additional packets, optimizing the network's defense over time for all flows $\{F_1, \dots, F_N\}$. 
In the context of NIDS, this constitutes an offline RL problem: cybersecurity is a critical domain where trial-and-error learning approaches are not viable due to the high stakes involved. 
Therefore, we rely solely on a sequence of historical network data for learning without the possibility of real-time detection within the network environment. This approach ensures that our learning and optimization strategies are derived from established patterns and behaviors, minimizing the risk of being attacked while enhancing the system's intrusion detection capabilities.

\textbf{Offline RL Problem Formulation via Sequence Modeling. }
In our approach, we adopt an alternative offline RL paradigm known as offline RL via sequence modeling for packet-level NIDS: rather than learning values of packet-response pairs and deriving a policy guided by these values or optimizing a policy with policy gradients, our objective is to directly map network security states to response decisions by uncovering the intrinsic relationships among security states, network responses, and packet characteristics. Consequently, an RL problem in this context is transformed into a supervised learning problem, more specifically, one of sequence modeling. This methodology allows for a more direct and nuanced understanding of the sequential nature of network traffic and its implications for security measures.

To streamline our discussion, we modify our notation system from the previously introduced \(\{(t_i, p_i, d_i, w_i)\}_{i=1}^{I_{n}}\) to \(\{(o_i, a_i)\}_{i=1}^{I_{n}}\), where \(o\) represents observables or observations containing packet characteristics and their inspection times, and \(a\) signifies actions encompassing the decisions and waiting periods. We conceptualize the inter-packet rewards \(r_i := \int_{t_i}^{t_{i+1}} s(t)dt\) as part of the data stream: \((r_1, o_1, a_1, ..., r_i, o_i, a_i, ...)\), and construct a sequence model $\Phi$ parameterized by \(\mu\) to model these data trajectories. The model, by observing the rewards \(r_i\), the current observation \(o_i\), and all previous history \(\mathcal{T}(i-1) := \{(r_j, o_j, a_j)\}_{j=1}^{i-1}\), can autoregressively yield appropriate response decisions. Thus, the training objective for each flow $F_n$ becomes one of density maximization: 
\begin{equation}\label{eq:2}
    \max_{\mu} \prod_{i=1}^{I_{n}} \Phi_\mu(a_i | r_i, o_i, \mathcal{T}(i - 1)).
\end{equation}
In scenarios where actions are discrete, the densities in the objective simply transform into probability masses. This adaptation allows the RL paradigm to transition smoothly between discrete and continuous action spaces. A prime example of RL via sequence modeling is the DT~\cite{dt}, which utilizes the robust transformer architecture~\cite{vaswani2017attention} and its extensive memory capacity to effectively model the reward-observation-action trajectory in network security contexts.

%% file: 5Methodology.tex
\section{Methodology}
\label{sec:method}

In this section, we adapt the Continuous-Time Transformer architecture in~\citet{chen2023contiformer} that elegantly handles temporal information, flexibly captures the continuous-time process, and is ideally suited for learning network security policies from offline network traffic data. We will first outline how the Decision Transformer adapts to learn a security-conditioned policy from offline data, then delve into the specifics of the model architecture, and ultimately discuss the training objectives.

\textbf{Preparing Packet Payload Features}
We present a method to extract and compress single-packet features to create more compact feature embeddings~\cite{chen2023explainable,chen2023ride}. First, we extract the packet feature from the raw packet capture (PCAP) files.
For each packet, $p_i$ in the testing sequence of packets to be $\xi = [p_1,p_2,\dots,p_I] \in \mathbb{R}^{I\times F}$,  where $I$ is the number of packet samples in the current flow and $F$ is the number of features in each sample, we used a packet extraction method introduced in~\citet{payloadbyte} to analyze PCAP files and extract relevant information. We structure the feature vector for each packet by capturing raw bytes from the packet data and extracting features from the packet header using the parser module $f_p$. To avoid potential overflow or truncation issues, a payload range of $N_p$ bytes was established, ensuring the incorporation of every byte. Each byte was then converted from its hex value to an integer ranging between 0 and 255, resulting in $N_p$ features. For cases with less than $N_p$ payload bytes, zero padding was applied to maintain a consistent feature vector structure. Thus, The raw payload $l_i$ in each packet $p_i$ is transformed to $\boldsymbol{X}_i = f_p(l_i)$, where $\mid \boldsymbol{X}_i\mid = N_p$. 
To label the extracted packets for the use of later classification training, we compared features from the PCAP files to ground truth flow-level data provided by each testing dataset. 
An identifier feature \emph{Flow ID} is established for each flow to give corresponding packets the connection statistics, facilitating subsequent packet joint embedding generation.
After obtaining packets $p_1,\dots,p_I$ with $N_p$ dimension transformed payload features $\boldsymbol{X}_i = [x^{1}_{i}, \dots, x^{N_p}_{i}]$, we train an autoencoder to learn a compressed representation of the payload features for each packet $p_i$. 
 Let $\boldsymbol{X}=[\boldsymbol{X}_1, \dots, \boldsymbol{X}_I] \in \mathbb{R}^{I \times N_p}$ be the input payloads from all packets $p_1,\dots,p_I$, where $I$ is the number of payload samples and $N_p$ is the number of payload features per sample. 
 The autoencoder consists of an encoder network, a decoder network, and a bottleneck layer with $N_b$ neurons, which is the number of payload features after compression. 
 The encoder network is defined as:
 \begin{equation} \label{eq:enc}
     \begin{aligned}
         \boldsymbol{H} = \sigma(W_1 \boldsymbol{X} + b_1),\ \ \boldsymbol{Z} = \sigma(W_2 \boldsymbol{H} + b_2),
     \end{aligned}
 \end{equation}
where $W_1 \in \mathbb{R}^{N_p \times h}$ and $W_2 \in \mathbb{R}^{h \times N_b}$ are the weight matrices of the encoder network, $b_1 \in \mathbb{R}^{h}$ and $b_2 \in \mathbb{R}^{N_b}$ are the bias vectors, $\sigma$ is the activation function (e.g., ReLU or sigmoid), and $\boldsymbol{H}$ is the hidden layer with $h$ neurons. The output of the encoder network is the compressed payload representation $\boldsymbol{Z} \in \mathbb{R}^{N_b}$.
The decoder network is defined as:
\begin{equation} \label{eq:dec}
     \begin{aligned}
         \boldsymbol{H}' = \sigma(W_3 z + b_3),\ \ \boldsymbol{X}' = \sigma(W_4 \boldsymbol{H}' + b_4),
     \end{aligned}
 \end{equation}
where $W_3 \in \mathbb{R}^{N_b \times h}$ and $W_4 \in \mathbb{R}^{h \times N_p}$ are the weight matrices of the decoder network, $b_3 \in \mathbb{R}^{h}$ and $b_4 \in \mathbb{R}^{N_p}$ are the bias vectors, $\sigma$ is the activation function, and $\boldsymbol{H}'$ is the hidden layer with $h$ neurons. The output of the decoder network is the reconstructed input data $\boldsymbol{X}' \in \mathbb{R}^{I \times N_p}$.
The objective of the autoencoder is to minimize the reconstruction error between the input data $X$ and the reconstructed output data $X'$ using a loss function, we use the mean squared error:
\begin{equation} \label{eq:ae_loss}
 L = \frac{1}{n} \sum_{k=1}^n | \boldsymbol{X}_k - \boldsymbol{X}_k'|^2,
\end{equation}
where $I$ is the number of payload samples, $\boldsymbol{X}_k$ is the original payload input data for the $k$-th sample, $\boldsymbol{X}_k'$ is the reconstructed output payload data for the $k$-th sample, and $|\cdot|$ denotes the L2-norm. 
In this case, once the autoencoder is fit, the reconstruction aspect of the model can be discarded and the model up to the point of the bottleneck can be used. The output $\boldsymbol{Z} \in \mathbb{R}^{N_b}$ of the model at the bottleneck is a fixed-length vector that provides a compressed representation of the input payload samples. In our experiments, we use the compressed payload features $\{z^{1}_{i}, \dots, z^{N_b}_{i}\}$ for packet characteristics $\{p_1,\dots,p_{I_n}\}$ for each flow $F_n, n\in \{1,\dots,N\}$.

\textbf{Trajectories Representation. }
We start by presenting our method, revisiting the continuous-time offline network intrusion detection RL objective Eq.~\eqref{eq:1}: at each inspection point \(i\) in the network, the aim is to discover a policy \(\pi\) that maximizes the subsequent cumulative network security condition, akin to the return-to-go in discrete-time RL. We refer to this as the return-to-go \( \hat{R}_i^\pi \): for each network check \(i\), based on decisions and wait times \((d_i, w_i)\) chosen according to policy \(\pi\), the objective is to maximize the expected cumulative security return, mathematically represented as:
\begin{equation}\label{eq:3}
  \hat{R}_i^{\pi} = \sum_{j=i}^{\infty} \underset{d_i,w_i \sim \pi}{\mathbb{E}}\left[\int_{t_j}^{t_{j+1}} s(t) dt\right] = \sum_{j=i}^{\infty}  \underset{d_i,w_i \sim \pi}{\mathbb{E}} r_j,  
\end{equation}
where \( r_j \) is the reward at each subsequent check.

In a discrete-time framework with evenly spaced intervals, rather than directly seeking a policy that maximizes the return-to-go \( \hat{R}_i^\pi \), the original Decision Transformer focuses on uncovering the relationships among the return-to-go’s \( \hat{R}_i \), observations \( p_i \), and actions \( d_i \) by representing them as a sequence of data in a trajectory form \( \tau = (\hat{R}_1, p_1, d_1, ..., \hat{R}_{I_{n}}, p_{I_{n}}, d_{I_{n}}) \). By setting an initial desired security return \( \hat{R}_{\text{initial}} \), the policy induced at each inspection point \( i \) is the action \( \hat{d}_i \) derived by autoregressively processing the trajectory data starting from \( (\hat{R}_{\text{initial}}, p_1) \). This approach allows the Decision Transformer to learn optimal responses sequentially based on past observations and desired future returns.
To adapt the Decision Transformer (DT) for continuous-time packet-level NID, we face two main challenges: 1) Modeling interval times \( w_i \) in a way that they lead to the desired security return \( \hat{R}_{\text{initial}} \), and 2) Adjusting the DT's attention mechanism to reflect the actual temporal distances in the irregular interval times.

Addressing the first problem, we suggest modeling the interval times \( w_i \) not just as another aspect of the decision-making process but also as giving them distinct tokens in the sequence model. This allows for a more nuanced capture of the interplay among network observations, response decisions, cumulative security conditions, and interval times. For instance, in the network context, the decision to delay the next packet inspection \( w_i \) is typically based on the current network state and response decision \( d_i \). Hence, we naturally consider a trajectory representation consisting of tuples (return-to-go, observation, decision, interval time): 
\begin{equation}\label{eq:4}
    \tau_{\text{modified}} = (\hat{R}_1, p_1, d_1, w_1, ..., \hat{R}_{I_{n}}, p_{I_{n}}, d_{I_{n}}, w_{I_{n}}),
\end{equation}
the training objective in Eq.~\eqref{eq:2} is then factorized as:
\begin{equation}\label{eq:5}
    \prod_{i=1}^{I_{n}} \left[\Phi_\mu(d_i | r_i, o_i, \mathcal{T}(i - 1)) \cdot \Phi_\mu(w_i | d_i, r_i, o_i, \mathcal{T}(i - 1))\right] .
\end{equation}
This trajectory representation and factorization are tailored to our specific application; the dependence of response decisions on interval times can be altered as needed.

To tackle the second challenge, we adopt the modifications to the continuous transformer architecture~\cite{chen2023contiformer}, which is designed to align the model's attention mechanism with the continuous and irregular nature of the packet inspection intervals, ensuring the model's relevance and efficacy in the context of packet-level NIDS.

\textbf{Model Architecture. }
To accommodate the irregular temporal intervals between network inspections in our packet-level NIDS, we implement a temporal position embedding in DT, enabling it to vary fluidly with time. 
We first denote each element over the total length of \( 4I_{n} \) (four elements per inspection \( i \)) in the modified trajectory representation \( \tau_{\text{modified}} \) defined in Eq.~\ref{eq:4} having a value $z_{\Omega}(t)$ at time $t$, where $\Omega \in \{\hat{R}, p, d, w\}$. $z_{\Omega}(t)$ could be a scalar (like a security score) or a vector (like multi-dimensional network states), which signifies the type of element (return-to-go, observation, decision, interval time).
For each element $z_{\Omega_j}(t_j)$ where \( j \in \{1, ..., 4I_{n}\} \), we tokenize it by separately embedding its temporal information \( t_j \), value \( z_j \), and type \( \Omega_j \). 

For embedding $t_j$, we adopt the sinusoidal temporal embedding methods~\cite{zuo2020transformer,yang2021transformer,chen2023contiformer}. For the \( k^{th} \) dimension of the temporal embedding, where \( k \in \{1, ..., d_{\text{time}}\}\), the embedding is given by \( \sin\left(\frac{t}{C_{}^{\frac{k}{d_{\text{time}}}}}\right) \) for even \( k \) and \( \cos\left(\frac{t}{C_{}^{\frac{k-1}{d_{\text{time}}}}}\right) \) for odd \( k \). Here we use \( C = 10000 \). The initial value embeddings \( z_j \) and type embeddings \( \Omega_j \) are derived through linear transformations. Instead of adding these embeddings, we opt for concatenation—combining the temporal embeddings, type embeddings, and value embeddings to form the base layer input tokens \( \text{Emb}^{(0)}:= [t_j; z_j; \Omega_j] \). This method ensures more direct access to the temporal information.

At each attention layer \( l \), the keys, queries, and values are derived as \( \psi_j^{(l)} = \Psi^{(l)}(\text{Emb}_{j}^{(l-1)}) \), where \( \psi \) represents either \( k \), \( q \), or \( v \) (key, query, value), and \( \Psi \) corresponds to their respective linear transformations \( K \), \( Q \), or \( V \). A causal mask is applied to the transformer, ensuring that each element \( z_{\Omega_j}(t_j) \) in \( \tau_{\text{modified}} \) with order index \( j \) attends only to preceding elements and itself in the sequence \( \{z_{\Omega_b}(t_b)\}_{b=1}^j \). The unnormalized attention weight assigned to an element \( z_{\Omega_b}(t_b) \) is denoted as \( \alpha(z_{\Omega_j}(t_j), z_{\Omega_b}(t_b))^{(l)} = q_{j}^{(l)} \cdot k_{b}^{(l)} \).
Finally, to obtain the embeddings for each layer \( \text{Emb}_{j}^{(l)} \), we first calculate:
\begin{equation}
    \sum_{b=1}^{j} \text{softmax} \left( \left[ \alpha(z_{\Omega_j}(t_j), y_{e_{b'}}(t_{b'}))^{(l)} \right]_{b'=1}^j \right)_{b}\cdot v_b^{(l)},
\end{equation}
and followed by layer normalization~\cite{ba2016layer}, a feed-forward connection, and a residual connection~\cite{radford2018improving,dt,chen2023contiformer}. 

\textbf{Training. }
During the training phase for our Decision Transformer in continuous-time packet-level NID, we compute the return-to-go \( \hat{R}_i = \sum_{k=i}^{I_n} r_k \) and the interval times \( w_i = t_{i+1} - t_i \) from the offline data. We then format these data into trajectories as per Eq. (3) for input into our transformer. In this study, we opt for deterministic prediction of both response decisions \( d_i \) and interval times \( w_i \). After processing through the transformer's last layer, we use additional fully connected network layers to directly output action weights estimates of the decisions \( \hat{d}_i \) and interval wait times \( \hat{w}_i \), then we apply $\mathrm{argmax}$ for outputting the discrete actions, e.g., determining the type of the network attack. Note that our architecture can also be used for continuous actions (e.g., degree of security measure adjustment) by directly outputting the decisions \( \hat{d}_i \) and interval wait times \( \hat{w}_i \).
Accordingly, our training objective is replaced with deterministic supervised learning losses. We employ cross-entropy loss for discrete actions (different types of attacks):
\begin{equation} \label{eq:loss_discrete}
    L_{\text{train}} = -\sum_{n=1}^{N}\sum_{i=1}^{I_n} \sum_{k=1}^{\text{\#types}} \log P(\hat{d}_i = k) \boldsymbol{1}\{d_i=k\},
\end{equation}
where \( \hat{d}_i \) represents the predicted decision (action type) for the \( i^{th} \) instance in the \( n^{th} \) flow data batch, \( d_i \) is the true decision (actual action type) for that instance, \( P(\hat{d}_i = k) \) is the probability assigned by the model that the \( i^{th} \) instance belongs to attack class \( k \), \( \boldsymbol{1}\{d_i=k\} \) is an indicator function that equals 1 if the true class of the \( i^{th} \) instance is \( k \), and 0 otherwise. The summation over \( N \) and \( I_n \) indicates that the loss is calculated over all packet samples in all flow batches, and the summation over \( \text{\#types} \) accounts for all possible types of network attacks.
The cross-entropy loss function thus penalizes the model when it assigns low probabilities to the true class of each instance, encouraging the model to improve its accuracy in classifying different types of network attacks.
For other NID problems with continuous actions, our method could also be used by changing the loss function to mean squared error $L_{\text{train}} = \frac{1}{I_n \cdot N} \sum_{n=1}^{N} \sum_{i=1}^{I_n} (\hat{d}_i - d_i)^2$. 
In both cases, the continuous interval times are trained with mean squared error.

During the evaluation phase of our continuous-time Decision Transformer for packet-level NIDS, we initialize the conditioning return-to-go \( \hat{R}_1 \) as a user-specified target security return. Following the approach in~\citet{dt,chen2023contiformer}, we typically select the highest return found in the offline dataset as a practical starting point. The subsequent conditioning returns are calculated as \( \hat{R}_i = \hat{R}_{i-1} - r_{i-1} \), deducting the actual security reward obtained at each inspection from the previous conditioning return. 
The decision for each packet inspection is then autoregressively determined based on the entire trajectory history, along with the current conditioning return-to-go and the observed network state at that point. Furthermore, the interval time for the next packet inspection is decided based on the current response decision. This process ensures that each step's decision and timing are informed by the most up-to-date and relevant network data, aligning with the overall objective of maintaining optimal network security.

%% file: 6Experiment.tex
\input{0table3}

\section{Experiments}


In this section, we present a framework to abstract network intrusion detection as a sequence modeling problem for decision-making and present an empirical study of applying the decision transformer. We evaluate our proposed decision transformer design together with several baseline algorithms on the first sequential decision-making-based malicious packet detection environment based on the UNSW-NB15~\cite{unsw} offline packet-level dataset through supervised learning as a normal classification job. We utilize the evaluated dataset at the packet level, derived from two distinct sources~\cite{chen2023ride,chen2023explainable}. These packets have been organized into flows using flow ID and timestamp criteria. Additionally, we've employed autoencoders to compress the payload characteristics of each packet, ensuring more efficient and compact representative embeddings for each packet's information.  Through the comparison of the two, we demonstrate the effectiveness of applying the decision transformer to the malicious packet detection problem in terms of both accuracy and timeliness. Furthermore, we analyze some critical properties of this problem setting to confirm the rationality of our motivation.

\textbf{Problem Formulation. }
In our network intrusion detection framework, the goal is to sequentially decide on actions for incoming network packets. We define the observation \(o_i(t)\) as the compact payload feature vector and packet information such as source IP, destination IP, source port, destination port, and protocol type, representing attributes of the \(i^{th}\) packet at time \(t\) in each flow $F_n, n\in \{1,\dots,N\}$, as detailed in Sections~\ref{sec:problem_formulation} and~\ref{sec:method}. The decision action space \(\mathcal{A} = \{0, 1, 2\}\) consists of three actions for decisions $d_i$: \(0\) for flagging a packet as benign, \(1\) for malicious, and \(2\) for waiting for more packets.
The packet reward function \(R_i(t)\) at time \(t\) is defined to reflect the accuracy of these actions:
\begin{equation} 
    R_i(t)_{(d_i(t), o_i(t))} = \begin{cases}
c_{\text{TP}} & \text{if } d_i(t) = 0 \text{ and } \text{L}(o_i(t)) = 0, \\
c_{\text{TN}} & \text{if } d_i(t) = 1 \text{ and } \text{L}(o_i(t)) = 1, \\
c_{\text{FP}} & \text{if } d_i(t) = 1 \text{ and } \text{L}(o_i(t)) = 0, \\
c_{\text{FN}} & \text{if } d_i(t) = 0 \text{ and } \text{L}(o_i(t)) = 1, \\
c_{\text{wait}} & \text{if } d_i(t) = 2.
\end{cases}
\end{equation}
Here, \(c_{\text{TP}}\), \(c_{\text{TN}}\), \(c_{\text{FP}}\), and \(c_{\text{FN}}\) represent the rewards for true positives, true negatives, false positives, and false negatives, respectively. \(c_{\text{wait}}\) is the reward for waiting. By adjusting these constants, we can tailor the reward system to prioritize either faster detection or higher accuracy, thus aligning the incentives with our specific detection objectives in NID. The true label of the packet \(i\) at time \(t\) is denoted by \(\text{L}(o_i(t))\). Agents responsible for each packet decision aim to minimize the cross-entropy loss defined in Eq.~\ref{eq:loss_discrete}. 
To measure decision \textbf{timeliness} in each flow, we define the \textbf{Time To Resolution (TTR)} as \( \frac{w_i}{w_n} \). This is the percentage of maximum wait time (\( w_n = t_{I_n} - t_1 \)) used to make a decision on packets in flow \( F_n \) which contains \( I_n \) packets.

\textbf{Experiment Setup. }
Given the absence of established benchmarks for sequential decision-making at the packet level in NID, we sample and generate the trajectories of the first reward-guided offline dataset for sequential decision-making and deep reinforcement learning from the UNSW-NB15 packet-level dataset, using 3 different \textbf{sampling policies} to illustrate the quality of the dataset and distinguish the performance of the testing algorithm, namely: \textbf{Expert}, \textbf{Medium}, and \textbf{Random}. Specifically, Expert dataset trajectories are sampled by simulating a policy that is able to make the decision with an overall accuracy of 90\% within the first 50\% length of the total packets, medium dataset trajectories are sampled with an overall accuracy of 50\% before the end of the last packet and the Random dataset trajectories are generated from a pure random walk. 

Each packet contains $N_b=100$ compressed payload bytes and additional packet information: source IP, destination IP, source port, destination port, and protocol type, along with the true label of the group of these packets being malicious or benign. We split the dataset into a training dataset and a testing dataset, where we sample the trajectories from the training dataset and evaluate the baseline policies on the testing dataset.
To make a relatively balanced dataset out of this newly generated packet-level dataset 
, we employ oversampling techniques on the malicious packets to ensure a more equitable representation of both benign and malicious traffic in our dataset. In this formulation, we also aim to strike a balance between the timeliness of decision-making and the accuracy of intrusion detection, this is achieved by two key factors during the evaluation as we consider the accuracy: the traditional and important metric, and reward: a metric used to quantify the timeliness of decision-making under a reinforcement learning problem formulation.


\textbf{Selected Baselines. }
This is the first work mainly for demonstrating a framework to abstract network intrusion detection as a sequence modeling problem for decision-making and testing the proposed algorithm on the first proposed sequential decision-making-based malicious packet detection environment
, we present preliminary results of comparing \textbf{Decision Transformer (DT)} with 3 different algorithms: reward-conditioned \textbf{behavior cloning (BC)}, an RL agent trained with \textbf{Conservative Q-Learning (CQL)}, and a 4-layer \textbf{Deep Neural Network classifier (DNN)} trained under supervised learning.

\textbf{General Performance. }
We evaluate and compare the performance of the proposed method with all selected baselines of the \textbf{accuracy} and \textbf{timeliness} of their policy learned. Specifically, we evaluate their performance in terms of several key metrics, where Accuracy, Precision, Recall, and F1-Score are used to evaluate the performance of it as a classification job; while reward (normalized score with 100 being expert policy and 0 being random policy) and resolution time (Time To Resolution) are used to quantify the timeliness of the decision making. All results are averaged over 3 different runs.


The results in Table~\ref{tab:my-table} demonstrate that our method outperforms all baselines in both accuracy and timeliness. With the \textbf{Expert} sampling policy, both DT and CQL achieve a Precision of 0.99, but DT records a lower Time To Resolution (TTR), indicating its ability to detect malicious packets earlier. In the \textbf{Medium} sampling scenario, DT and BC both achieve a TTR of 0.80, yet DT shows superior detection accuracy and reward, proving its effectiveness even with less ideal training data. Under \textbf{Random} sampling, DT stands out by exceeding a 0.91 detection accuracy score while maintaining the lowest TTR, showcasing its robustness in learning from non-optimal datasets. This highlights DT's potential in practical NID situations, especially in detecting malicious activities early, which is crucial for tasks with more strict security requirements and limited non-optimal training data, we can still make more swift detections with acceptable accuracy scores under limited training resources.





%% file: 0table3.tex
\begin{table*}[ht] 
\centering
\resizebox{0.69\textwidth}{!}{%
\begin{tabular}{@{}lccccccc@{}}
\toprule
Dataset &
  \textbf{Model} &
  \textbf{Accuracy(\%)} &
  \textbf{Precision} &
  \textbf{F1-Score} &
  \textbf{Recall} &
  \multicolumn{1}{l}{\textbf{Reward}} &
  \multicolumn{1}{l}{\textbf{TTR}} \\ \midrule
\multicolumn{1}{c}{\multirow{4}{*}{Expert}} 
                        & DT  & \textbf{99.33} &  \textbf{0.99} &  \textbf{0.99} &  \textbf{0.99} &  \textbf{81.4} &  \textbf{0.81} \\
\multicolumn{1}{c}{}    & BC  & 96.27 & 0.94 & 0.96 & 0.98 & 58.8 & 0.86\\
\multicolumn{1}{c}{}    & CQL & 97.82 & \textbf{0.99} & 0.97 & 0.96 & 61.3 & 0.85\\
\multicolumn{1}{c}{}    & DNN & 88.30 & 0.82 & 0.89 & 0.97 & - & - \\ \midrule

\multirow{4}{*}{Medium} & DT  & \textbf{98.84} & \textbf{0.99} & \textbf{0.98} & \textbf{0.98} & \textbf{64.3} & \textbf{0.80} \\
                        & BC  & 95.34 & 0.96 & 0.95 & 0.95 & 42.1 & \textbf{0.80} \\
                        & CQL & 96.09 & 0.97 & 0.96 & 0.94 & 43.3 & 0.89 \\
                        & DNN & 86.54 & 0.79 & 0.88 & 0.99 & - & - \\ \midrule
                        
\multirow{4}{*}{Random} & DT  & \textbf{94.19} & \textbf{0.91} & \textbf{0.94} & \textbf{0.97} & \textbf{40.4} & \textbf{0.91} \\
                        & BC  & 50.11 & 0.27 & 0.01 & 0.06 & 33.2 & 0.92 \\
                        & CQL & 66.40 & 0.55 & 0.71 & 0.96 & 39.2 & \textbf{0.91} \\
                        & DNN & 49.31 & 0.10 & 0.01 & 0.01 & - & - \\ \bottomrule
\end{tabular}%
}
\caption{Decision Transformer (DT) achieves the highest detection accuracy scores and normalized reward under all sampling policies, while also achieving the lowest decision time as measured by the Time To Resolution (TTR).}
\label{tab:my-table}
\vspace{-0.1in}
\end{table*}


%% file: 7Conclusion.tex
\section{Conclusion and Future Work}

In this paper, we propose a new framework for network intrusion detection by formulating it as a sequential decision-making problem and applying offline reinforcement learning techniques. We adopt the Decision Transformer architecture to model the continuous network traffic data for timely packet-level detection. 
A key contribution is the introduction of the tradeoff between detection accuracy and timeliness using a reward function that accounts for both correct and fast threat identification.
Experiments on data derived from the UNSW-NB15 dataset show our method balances response speed and precision, outperforming baselines in accuracy, precision, recall, and F1-score while achieving higher rewards that quantify timelier detection. 
The results validate the capability of modeling the sequential decision process using decision transformers as an offline RL problem, opening research avenues into specialized architectures and embeddings for this application.
These results also open the door for future explorations in translating these algorithms to dedicated hardware that can support real-time performance operation at the edge. For example, emerging technologies have shown significant promise for efficient implementation of transformer networks {~\cite{memristortransformer}}. Future work will explore the software/hardware tradeoffs of decision transformers under additional resource constraints such as area and power consumption.